\documentclass[pra,twocolumn,tightenlines,showpacs,nofootinbib]{revtex4}
\usepackage{multirow}
\usepackage{amsmath}
\usepackage{bm}
\usepackage{dcolumn}

\begin{document}
\title{Polarizabilities of Si$^{2+}$: a benchmark test of theory and experiment}
\author{M. S. Safronova$^1$}
\author{S. G. Porsev$^{1,2}$}
\author{M. G. Kozlov$^2$}
\author{Charles W. Clark$^3$}
\affiliation{$^1$Department of Physics and Astronomy, University of Delaware,
                  Newark, Delaware 19716, USA\\
$^2$Petersburg Nuclear Physics Institute, Gatchina,
                  Leningrad District, 188300, Russia\\
$^3$Joint Quantum Institute, National Institute of Standards and Technology and the
 \\University of Maryland, Gaithersburg, Maryland, 20899-8410,
USA}

\date{\today}

\begin{abstract}
We have calculated electric-dipole polarizabilities of the $3s^2~^1S_0$,
 $3s3p~^3P_0$, and $3s3p~^1P_1$ states of the Si$^{2+}$ ion using recently
developed configuration interaction + all-order method. Detailed evaluation of the uncertainties of the final results is carried out. Our value for
the ground state electric-dipole polarizability 11.670(13)~a.u. is in excellent agreement with the resonant excitation Stark ionization spectroscopy
value 11.669(9)~a.u. [Komara \textit{et al.}, J. Phys. B \textbf{38}, 87 (2005); Mitroy, Phys. Rev. A \textbf{78}, 052515 (2008)]. This work
represents the most precise benchmark test to date of theory and experiment in divalent atoms. The near cancellation of the $ns^2~^1S_0$ ground state
and the lowest $nsnp~^3P_0$ polarizabilities previously observed in B$^+$, Al$^+$, In$^+$, Tl$^+$, and Pb$^{2+}$ is also found in Si$^{2+}$ ion.
\end{abstract}
\maketitle

\section{Introduction}

The atomic dipole polarizability describes the first-order response
of an atom to an applied electric field. Atomic polarizabilities
have been the subject of considerable interest and heightened
importance in recent years
 due to a number of  applications, including  development of next-generation optical atomic clocks, optical cooling
and trapping schemes, quantum information with atoms and ions, tests of fundamental symmetries, studies of cold degenerate gases, thermometry and
other macroscopic standards,  study of long-range interactions, and atomic transition rate determinations~\cite{MitSafCla10}. An imperfect knowledge
of atomic polarizabilities is one of the the largest sources of uncertainty in the new generation of optical frequency standards
~\cite{MitSafCla10,SafJiaAro10}.

 Most of the applications listed above involve monovalent or divalent atoms and ions.
 There are a number of high-precision benchmark tests of
experimental and theoretical values for the polarizabilities of monovalent systems
\cite{EksSchCha95,DerJohSaf99,AmiGou03,SafCla04,AuzBluFer07,GunEllSaf07,IskSaf08,MitSaf09,MitSafCla10,HolRevLon10,KorTinGra11}. However, there are
few high-precision experimental data for the polarizabilities of  divalent systems which are of particular interest to optical clock
development~\cite{PorLudBoy08,RosHumSch08,JiaLudLem11,HacMiyPor08} and quantum information \cite{GorReyDal09}. Most recent data for polarizability
and Stark shifts of divalent systems are compiled in Tables 11, 13, and 14 of Ref.~\cite{MitSafCla10}.

For first row monovalent systems, such as Li and Be$^+$, the highest precision determination of the polarizabilities by theoretical and experimental
methods are found to be in good agreement  (see recent reviews~\cite{Mar06,MitSafCla10} and references therein). Here, we provide such a comparison
for a second row divalent species, Si$^{2+}$. We believe that the experimental data for this ion provides the most precise value of the
polarizability of any atomic system with two valence electrons. The theoretical calculation presented in the present work has the lowest uncertainty
reported for any divalent polarizability.

\subsection{Experimental determination of polarizabilities from Rydberg spectra}
The polarizability of an ion can  be extracted from the energies of the non-penetrating Rydberg series of the corresponding parent system (see
\cite{WarStuLun96} and references therein). The polarization interaction between the ionic core and the Rydberg electron shifts the energy levels
away from their hydrogenic values. If the Rydberg electron is in a high angular momentum state, it has negligible overlap with the core. In such
cases, the polarization interaction provides the dominant contribution to the energy shift. This effect is utilized in resonant excitation Stark
ionization spectroscopy (RESIS)
\cite{KomGeaFeh05,JacKomStu00,WarStuLun96,SnoLun08,KomGeaLun03,LunFeh07,SnoGeaKom05,HanKeeLun10,KeeHanWoo11,KeeLunFeh11}. $\textrm{RESIS}$
experiments have been extremely successful in high-precision determination of the ground state polarizabilities of H$^{+}_{2}$ and D$^{+}_{2}$
\cite{JacKomStu00}, Ne$^+$ \cite{WarStuLun96}, Na-like Mg$^{+}$ \cite{SnoLun08}, Na-like Si$^{3+}$ \cite{KomGeaLun03}, Mg-like
Si$^{2+}$~\cite{KomGeaFeh05}, Zn-like Kr$^{6+}$ \cite{LunFeh07}, Ba$^+$\cite{SnoGeaKom05}, Hg-like Pb$^{2+}$ \cite{HanKeeLun10}, Fr-like Th$^{3+}$
\cite{KeeHanWoo11}, and Rn-like Th$^{3+}$ \cite{KeeLunFeh11}. Quadrupole polarizabilities and transition matrix elements have also been determined
for some of these systems.

To the best of our knowledge, RESIS experiments provide the most
precise values known to date of
 the polarizability of any divalent atomic system.
  The most precise measurement has been carried out for the $3s^2~^1S_0$ ground state
of the Si$^{2+}$ ion; $\alpha_0=11.666(4)$~a.u~\cite{KomGeaFeh05}. Later analysis of the RESIS data that included additional terms in the
polarization expansion yielded $\alpha_0=11.669(9)$~a.u~\cite{Mit08}. Therefore, the Si$^{2+}$ RESIS experiment presents an excellent opportunity for
a high-precision benchmark comparison of theory and experiment.

 In this work, we use a recently
developed configuration iteration (CI)+ all-order method ~\cite{Koz04,SafKozJoh09,SafKozCla11}
 to calculate  properties of
Si$^{2+}$. Our value for the ground state electric-dipole
polarizability of 11.670(13)~a.u.
 is in excellent agreement with the RESIS result. Our
previous calculation of the Hg-like Pb$^{2+}$ ground state polarizability \cite{SafKozSaf12} was also in agreement with the RESIS value (accurate to
0.6\%)  within our estimated accuracy.

\begin{table*}
\caption{\label{table1}Comparison of experimental \cite{RalKraRea11} and theoretical energy levels in cm$^{-1}$. Two-electron binding energies are
given in the first row, energies in other rows are given relative to the ground state. Results of the CI, CI+MBPT, and CI+all-order calculations are
given in columns labeled CI, CI+MBPT, and CI+All. Corresponding relative differences of these three calculations with experiment are given in the
last three columns in \%.}
\begin{ruledtabular}
\begin{tabular}{lrccccccrrr}
\multicolumn{1}{c}{\multirow{2}{*}{State}} & \multicolumn{1}{c}{\multirow{2}{*}{Expt.}} & \multicolumn{1}{c}{\multirow{2}{*}{CI}} &
\multicolumn{1}{l}{\multirow{2}{*}{CI+MBPT}} &
 \multicolumn{1}{c}{\multirow{2}{*}{CI+All}}
& \multicolumn{3}{c}{Differences (cm$^{-1}$)}
 &\multicolumn{3}{c}{Differences (\%)} \\
&&&&&\multicolumn{1}{c}{CI} & \multicolumn{1}{c}{CI+MBPT} & \multicolumn{1}{c}{CI+All}& \multicolumn{1}{r}{CI} & \multicolumn{1}{r}{CI+MBPT} &
\multicolumn{1}{c}{CI+All} \\ \hline
$3s^2~^ 1S_0 $  &  634232 &  628511 &  634110 &  634226 &  $-$5722&   $-$123 & $-$7 &$-$0.9\% & $-$0.019\%& $-$0.001\%    \\
$3p^2~^ 1D_2 $  &  122215 &  120224 &  122225 &  122294 &  $-$1991&    10  & 80 &    $-$1.6\% &    0.008\% &   0.065\%      \\
$3p^2~^ 3P_0 $  &  129708 &  128589 &  129745 &  129753 &  $-$1119&    36  & 45 &    $-$0.9\% &    0.028\% &   0.035\%      \\
$3p^2~^ 3P_1 $  &  129842 &  128717 &  129878 &  129887 &  $-$1125&    36  & 45 &    $-$0.9\% &    0.028\% &   0.035\%      \\
$3p^2~^ 3P_2 $  &  130101 &  128964 &  130136 &  130145 &  $-$1137&    35  & 44 &    $-$0.9\% &    0.027\% &   0.034\%      \\
$3s3d~^ 3D_3 $  &  142944 &  141676 &  142953 &  142944 &  $-$1268&    10  & 1  &    $-$0.9\% &    0.007\% &   0.000\%      \\
$3s3d~^ 3D_2 $  &  142946 &  141678 &  142955 &  142946 &  $-$1267&    10  & 1  &    $-$0.9\% &    0.007\% &   0.000\%      \\
$3s3d~^ 3D_1 $  &  142948 &  141681 &  142957 &  142948 &  $-$1268&    9   & 0  &    $-$0.9\% &    0.006\% &   0.000\%      \\
$3s4s~^ 3S_1 $  &  153377 &  151756 &  153357 &  153403 &  $-$1621&   $-$20  & 26 &  $-$1.1\% & $-$0.013\%&    0.017\%      \\
$3p^2~ ^1S_0 $  &  153444 &  152674 &  153631 &  153613 &  $-$771 &    187 & 169&    $-$0.5\% &    0.122\% &   0.110\%      \\
$3s4s~ ^1S_0 $  &  159070 &  157543 &  159079 &  159116 &  $-$1527&    9   & 47 &    $-$1.0\% &    0.006\% &   0.029\%      \\
$3s3d~ ^1D_2 $  &  165765 &  165071 &  165937 &  165898 &  $-$694 &    172 & 133&    $-$0.4\% &    0.104\% &   0.080\%      \\  [0.5pc]
$3s3p~^ 3P_0 $  &  52725  &  51559  &  52722  &  52770  &  $-$1166&    $-$3  & 45 &  $-$2.2\% & $-$0.006\%&    0.086\%       \\
$3s3p~^ 3P_1 $  &  52853  &  51682  &  52849  &  52897  &  $-$1171&    $-$4  & 44 &  $-$2.2\% & $-$0.008\%&    0.083\%       \\
$3s3p~^ 3P_2 $  &  53115  &  51934  &  53110  &  53159  &  $-$1181&    $-$5  & 44 &  $-$2.2\% & $-$0.010\%&    0.082\%       \\
$3s3p~^ 1P_1 $  &  82884  &  82998  &  82969  &  82933  &      113&     84 & 48 &       0.1\% &    0.102\%&    0.058\%      \\
$3s4p~^ 3P_0 $  &  175230 &  173409 &  175202 &  175249 &  $-$1821&    $-$28 & 19 &  $-$1.0\% & $-$0.016\%&    0.011\%       \\
$3s4p~^ 3P_1 $  &  175263 &  173441 &  175235 &  175282 &  $-$1822&    $-$28 & 19 &  $-$1.0\% & $-$0.016\%&    0.011\%       \\
$3s4p~^ 3P_2 $  &  175336 &  173511 &  175308 &  175355 &  $-$1825&    $-$28 & 18 &  $-$1.0\% & $-$0.016\%&    0.011\%       \\
$3s4p~^ 1P_1 $  &  176487 &  174807 &  176469 &  176511 &  $-$1680&    $-$18 & 23 &  $-$1.0\% & $-$0.010\%&    0.013\%       \\
\end{tabular}
\end{ruledtabular}
\end{table*}
We note that Mg-like Si$^{2+}$ is a particularly interesting test
system due to its similarity with Mg-like Al$^+$, which was used to
construct an optical
 clock with a fractional frequency uncertainty of
 $8.6\times10^{-18}$~\cite{ChoHumKoe10}, the smallest such
 uncertainty yet attained.
 At room temperature, one of the largest contributions to the uncertainty
 budget of this clock is the blackbody radiation (BBR) shift.
The BBR frequency shift of a
  clock transition is related to the difference of the static electric-dipole
  polarizabilities between the  two clock states
\cite{PorDer06}. We have recently calculated this effect in Al$^+$ using the same CI+all-order approach. Excellent agrement of our present
calculation with the experiment in the ground state of Si$^{2+}$ provides an additional test of the  approach. At the present time, there are no
experimental data on the polarizabilities of the excited states of Si$^{2+}$ to the best of our knowledge.

\section{Method}

To evaluate uncertainties of the final results, we carry out three calculations in
different approximations: CI \cite{KotTup87}, CI+many-body
perturbation theory (MBPT) \cite{DzuFlaKoz96b}, and CI+all-order \cite{Koz04,SafKozJoh09,SafKozCla11}.
 These methods have been described in a number
of papers \cite{KotTup87,DzuFlaKoz96b,SafKozJoh09,SafKozCla11} and
 we provide only a brief outline of these approaches and a few details relevant to
this particular work.

Our point of departure is a solution of the Dirac-Fock  (DF) equations
 $$ \hat H_0\, \psi_c = \varepsilon_c \,\psi_c, $$
  where
$H_0$ is the relativistic DF Hamiltonian \cite{DzuFlaKoz96b,SafKozJoh09} and $\psi_c$ and $\varepsilon_c$ are single-electron wave functions and
energies. Self-consistent calculations were performed for the [$1s^2 2s^2 2p^6$] closed core, and  the $3s$, $3p$, $3d$, $4s$, $4p$, and $4d$
orbitals were formed in this potential. We constructed the B-spline basis
 set  consisting of $N=35$ orbitals for each of the $s,~ p_{1/2},~ p_{3/2},~ ...$ partial waves up to
$l\leq5$. The basis set is
formed in a spherical cavity with radius 60 a.u.
 The CI space is effectively complete and  includes $23$ orbitals for each partial wave with $l=0\ldots 4$.

The wave functions and the low-lying energy levels are determined by solving the multiparticle relativistic equation for two valence electrons
\cite{KotTup87}: $$ H_{\rm eff}(E_n) \Phi_n = E_n \Phi_n.$$
 The effective Hamiltonian is defined as $$ H_{\rm eff}(E) = H_{\rm FC} +
\Sigma(E),$$ where $H_{\rm FC}$ is the Hamiltonian in the frozen-core
approximation. The energy-dependent operator $\Sigma(E)$
 takes into account virtual core excitations. It is zero  in a pure CI calculation.
The  $\Sigma(E)$ part of the effective Hamiltonian is constructed
using second-order perturbation theory in the CI+MBPT approach
\cite{DzuFlaKoz96b} and linearized coupled-cluster single-double
method  in the CI+all-order approach \cite{SafKozJoh09}. Construction
of the effective Hamiltonian in the CI+MBPT and CI+all-order
approximations is described in detail in
Refs.~\cite{DzuFlaKoz96b,SafKozJoh09}. The dominant part of the Breit
interaction is included as described in Ref.~\cite{KozPorTup00}.

The scalar polarizability $\alpha_0$
 is separated into a valence polarizability $\alpha_0^v$, ionic core
polarizability $\alpha_c$, and a small term $\alpha_{vc}$
 that modifies ionic core polarizability due to
 the presence of two valence
electrons. The last two terms are  evaluated in the  random-phase approximation (RPA). Their uncertainty is determined by comparing the DF and RPA
values. The small $\alpha_{vc}$ term is calculated by adding vc contributions from the individual electrons, i.e.
$\alpha_{vc}(3s^2)=2\alpha_{vc}(3s)$, and $\alpha_{vc}(3s3p)=\alpha_{vc}(3s)+\alpha_{vc}(3p)$.

The valence part of the polarizability is determined  by solving the inhomogeneous
 equation in valence space, which is approximated  as
\cite{KozPor99a}
\begin{equation}
(E_v - H_{\textrm{eff}})|\Psi(v,M^{\prime})\rangle = D_{\textrm{eff}} |\Psi_0(v,J,M)\rangle
\end{equation}
for the state  $v$ with total angular momentum $J$ and projection $M$. The wave function $\Psi(v,M^{\prime})$ is composed of parts that have angular
momenta of $J^{\prime}=J,J \pm 1$ that allows us to determine the scalar and tensor polarizability of the state $|v,J,M\rangle$  \cite{KozPor99a}.
The effective dipole operator $D_{\textrm{eff}}$ includes RPA corrections.

 Unless stated otherwise, we use atomic units (a.u.) for all matrix
elements and polarizabilities throughout this paper: the numerical values of the
elementary charge, $e$, the reduced Planck constant, $\hbar = h/2
\pi$, and the electron mass, $m_e$, are set equal to 1. The atomic unit for
polarizability can be converted to SI units via
$\alpha/h$~[Hz/(V/m)$^2$]=2.48832$\times10^{-8}\alpha$~(a.u.), where the conversion
coefficient is $4\pi \epsilon_0 a^3_0/h$ and the Planck constant
$h$ is factored out in order to provide direct conversion into frequency units;
$a_0$ is the Bohr radius and $\epsilon_0$ is the electric constant.

\section{Results}
Comparison of the energy levels (in cm$^{-1}$) obtained in the CI,
CI+MBPT, and CI+all-order approximations with experimental values
\cite{RalKraRea11} is given in Table~\ref{table1}. Corresponding
relative differences of these three calculations from experiment are
given in the last three columns. Two-electron binding energies are
given in the first row of Table~\ref{table1}, energies in other rows
are measured from the ground state. For a few of the levels, the
accuracy of the CI+MBPT calculation is already on the order of our
expected precision. The accuracy of the ground state two-electron
binding energy is significantly improved in the CI+all-order
calculation in comparison with the CI+MBPT one; the CI+MBPT value
differs from the experiment by $-$123~cm$^{-1}$, while our all-order
value differs from the experiment by only $-$7~cm$^{-1}$ (see line
one of Table~\ref{table1}). The inclusion of the all-order
core-valence correlations significantly improves the differences
between the singlet and triplet states. For example, the CI+all-order
value of the $3s3p~^1P_1 - 3s3p~^3P_1$ interval, 30035~cm$^{-1}$,
differs by only 4~cm$^{-1}$ from the experimental value
30031~cm$^{-1}$. The corresponding CI+MBPT value, 30120~cm$^{-1}$,
differs from the experiment by 89~cm$^{-1}$. As a result, the
accuracy of the transition energies used in the polarizability
calculations improves in the  CI+all-order approach.

 We separated the effect of the Breit interaction by comparing the results of the calculations with
and without the Breit. The Breit contribution to the energies is very small, 0.01\% or less.
 However, the inclusion of the Breit interaction
significantly  improves the splittings of all triplet states. For example, the
$3s3p~^3P_1 - 3s3p~^3P_0$ and $3s3p~^3P_2 - 3s3p~^3P_0$ splittings are
136~cm$^{-1}$ and 413~cm$^{-1}$ without Breit, respectively.  The values of these
splittings in our final calculations that include Breit, are
128~cm$^{-1}$ and 389~cm$^{-1}$, in excellent agreement with  the experimental values,
129~cm$^{-1}$ and 390~cm$^{-1}$

 We note that the transition energies relevant to the calculations of the $3s3p~^3P_0$
 polarizabilities are more accurate than
the energies relative to the ground state listed in Table~\ref{table1}.

  \begin{table*}
\caption{\label{table2}Contributions to the  $3s^2\;^1S_0$, $3s3p\;^3P_0$, and  $3s3p\;^1P_1$ polarizabilities of Si$^{2+}$ in a.u. The dominant
contributions to the valence polarizabilities are listed separately with the corresponding absolute values of electric-dipole reduced matrix elements
given in columns labeled $D$. The theoretical and experimental \cite{RalKraRea11} transition energies are given in columns $\Delta E_{\rm th}$  and
$\Delta E_{\rm expt}$. The remaining contributions to valence polarizability are given in rows Other. The contributions from the core and vc terms
are given in rows $\alpha_c$ and $\alpha_{vc}$, respectively. The dominant contributions to $\alpha_0$ listed in columns $\alpha_0(\mathrm{A})$ and
$\alpha_0(\mathrm{B})$ are calculated with CI + all-order and experimental energies \cite{RalKraRea11}, respectively.}
\begin{ruledtabular}
\begin{tabular}{llrrcrr}
  \multicolumn{1}{l}{State} &\multicolumn{1}{l}{Contribution} & \multicolumn{1}{c}{$\Delta E_{\rm expt}$} & \multicolumn{1}{c}{$\Delta E_{\rm th}$} & \multicolumn{1}{c}{$D$} &
\multicolumn{1}{c}{$\alpha_0(\mathrm{A})$} &\multicolumn{1}{c}{$\alpha_0(\mathrm{B})$}
\\
\hline
$3s^2\; ^1S_0$&$3s^2\; ^1S_0 - 3s3p\; ^1P_1$  & 82884  & 82933  & 2.539 &  11.375 & 11.382   \\
&$3s^2\; ^1S_0 - 3s4p\; ^1P_1$                & 176487 & 176511 & 0.198 &  0.032  & 0.032   \\
&Other                                        &        &        &       &  0.105  & 0.105   \\
&$\alpha_c$                                   &        &        &       &  0.162  & 0.162   \\
&$\alpha_{vc}$                                &        &        &       &  $-$0.011 & $-$0.011  \\
&Total                                        &        &        &       &   11.664&  11.670 \\
[0.5pc]
$3s3p\; ^3P_0$&$3s3p\; ^3P_0 - 3p^2~^3P_1$    &  77117 &  77117 &  1.516 &  4.359  & 4.359     \\
&$3s3p\; ^3P_0 - 3s3d~^3D_1$                  &  90224 &  90179 &  1.779 &  5.137  & 5.135    \\
&$3s3p\; ^3P_0 - 3s4s~^3S_1$                  &  100652&  100633&  0.628 &  0.573  & 0.573    \\
&Other                                        &        &        &        &  0.201  & 0.201    \\
&$\alpha_c$                                   &        &        &        &  0.162  & 0.162    \\
&$\alpha_{vc}$                                &        &        &        &  $-$0.006 & $-$0.006   \\
&Total                                        &        &        &        &  10.427 & 10.425   \\
[0.5pc]
$3s3p\; ^1P_1$&$3s3p\; ^1P_1 - 3s^2~ ^1S_0$&   $-$82884&  $-$82933 & 2.539&   $-$3.792 & $-$3.794      \\
              &$3s3p\; ^1P_1 - 3p^2~ ^1D_2$&   39330 &  39361  & 1.074&   1.428  & 1.429      \\
              &$3s3p\; ^1P_1 - 3p^2~ ^1S_0$&   70560 &  70680  & 1.776&   2.178  & 2.181      \\
              &$3s3p\; ^1P_1 - 3s4s~ ^1S_0$&   76185 &  76184  & 0.996&   0.634  & 0.634      \\
              &$3s3p\; ^1P_1 - 3s3d~ ^1D_2$&   82881 &  82965  & 4.450&   11.642 & 11.654     \\
 &   Other                                 &         &         &      &   0.440  & 0.440      \\
&$\alpha_c$                                &         &         &      &    0.162 &  0.162     \\
&$\alpha_{vc}$                             &         &         &      &    $-$0.006&  $-$0.006    \\
&Total                                     &         &         &      &    12.686&  12.701    \\
\end{tabular}
\end{ruledtabular}
\end{table*}

 \begin{table}
\caption{\label{table3} Summary of the results for the  $3s^2\;^1S_0$, $3s3p\;^3P_0$, and  $3s3p\;^1P_1$ polarizabilities of Si$^{2+}$ in a.u. and
the evaluation of the uncertainties. First three rows give \textit{ab initio} results for valence polarizabilities calculated in the CI, CI+MBPT, and
CI+all-order approximations.  In the CI+All (B) calculation, theoretical energies are replaced by the experimental values for the dominant
contributions. The final results listed in row  Total $\alpha_0 $  are compared with other theory~\cite{Mit08} and
experiment~\cite{KomGeaFeh05,Mit08}.}
\begin{ruledtabular}
\begin{tabular}{lrrr}
\multicolumn{1}{l}{Method} & \multicolumn{1}{r}{$3s^2~^1S_0$} & \multicolumn{1}{r}{$3s3p~^3P_0$} & \multicolumn{1}{r}{$3s3p~^1P_1$}\\
\hline
 CI (A)                      &11.567 & 10.353 & 13.040  \\
 CI+MBPT (A)                 &11.502 & 10.262 & 12.539 \\
 CI+All (A)                  &11.512 & 10.271 & 12.530 \\
 CI+All (B)                  &11.519 & 10.268 & 12.545  \\
 Diff. All $-$ MBPT           &0.010  & 0.009  & $-$0.009  \\
 Diff. (B)$-$(A)               &0.007  & $-$0.003 & 0.015  \\         \hline
Final  $\alpha^{v}_0$       &11.519(10) & 10.268(9) & 12.545(15)  \\
$\alpha_{c}$                                              &    0.162(9) &   0.162(9) &   0.162(9)\\
$\alpha_{vc}$                                        &  $-$0.011(2) &  $-$0.006(1) &  $-$0.006(1)    \\
Total $\alpha_0 $   &            11.670(13) & 10.425(13) & 12.701(17)                \\
Theory \cite{Mit08}             & 11.688  &&    12.707       \\
Theory \cite{HamHib08} &          11.75        &&  \\
Expt. \cite{KomGeaFeh05}    &                          11.666(4) &&  \\
Expt. \cite{KomGeaFeh05,Mit08}\footnotemark    &                          11.669(9) &&  \\
\end{tabular}
\end{ruledtabular}
\footnotetext{This value is a result of revised analysis \cite{Mit08} of the RESIS experiment  \cite{KomGeaFeh05}.}
\end{table}

While we do not use the sum-over-state approach in the calculation of the polarizabilities, it is useful to establish
 the dominant contributions to the final values. We
combine our CI+all-order results for the electric-dipole matrix
elements and energies  according to the sum-over-states formula for
the valence polarizability~\cite{MitSafCla10}:
\begin{equation}\label{genpol}
\alpha_0^v = \frac{2}{3(2J+1)}\sum_n\frac{|\langle v\| D\| n\rangle|^2}{E_n-E_v}
\end{equation}
to calculate the contribution of specific transitions.
 Here,  $J$ is the total angular momentum of the state $v$, $D$ is the electric-dipole operator, and $E_i$ is the energy of the state $i$.
The breakdown of the contributions to the $3s^2\ ^1S_0$, $3s3p\ ^3P_0$, and $3s3p\ ^1P_1$ scalar  polarizabilities  $\alpha _0$ of Si$^{2+}$ in a.u.
is given in Table~\ref{table2}. Absolute values of the corresponding reduced electric-dipole matrix elements are listed in column labeled ``$D$'' in
$a_{0}e$. The theoretical and experimental   \cite{RalKraRea11} transition energies  are given in columns $\Delta E_{\rm th}$  and $\Delta E_{\rm
expt}$. The remaining valence contributions are given in rows Other. The contributions from the core and vc terms are listed in rows  $\alpha_c$ and
$\alpha_{vc}$, respectively. The dominant contributions to $\alpha_0$ listed in columns $\alpha_0({\rm A})$ and $\alpha_0({\rm B})$ are calculated
with CI + all-order energies and experimental \cite{RalKraRea11} energies, respectively.
 The differences between $\alpha_0({\rm A})$ and  $\alpha_0({\rm B})$ values are small  due to excellent agreement of the corresponding transition energies with
experiment. We take $\alpha_0({\rm B})$ results as final. Our study of the  Breit interaction shows that it contributes only 0.03-0.07\% to the
\textit{ab initio} values of  polarizabilities.

\section{Evaluation of the uncertainty and conclusion}

There are three contributions to the uncertainties in the final polarizability values  that  arise from the uncertainties in the valence
$\alpha^v_0$, core $\alpha_c$, and vc $\alpha_{vc}$ polarizability terms. To evaluate uncertainty in the valence polarizabilities, we compare the
results of the CI, CI+MBPT, CI+all-order calculations with our final CI+all-order calculation in which energies in the dominant contributions are
replaced by their experimental values. The results of the last two calculations are given in Table~\ref{table2} in columns $\alpha_0(\mathrm{A})$ and
$\alpha_0(\mathrm{B})$. We summarize the results of all four calculations in Table~\ref{table3}. For consistency, we refer to these calculations as
CI~(A), CI+MBPT~(A), CI+All~(A), and CI+All~(B) since only theoretical energies (in the corresponding approximation) were used in the first three
calculations. We evaluate the uncertainty of the final results in two different ways: (1) as the difference between the CI+all-order and CI+MBPT
calculations, listed in row labeled Diff.~(All-MBPT), and (2) as the difference between the CI+all-order results with theoretical and experimental
energies, listed in row labeled Diff.~(B)-(A). We take the \textit{largest} of the two uncertainties as the final uncertainty in the valence
polarizability $\alpha_0^v$. The uncertainty analysis is carried out separately for each state.

 To evaluate the
uncertainty in the $\alpha_{c}$ and $\alpha_{vc}$ contribution to
the polarizability, we calculate these terms in both DF and RPA
approximations. The DF values for the $\alpha_{c}$ and
$\alpha_{vc}(3s^2)$ are 0.153~a.u. and -0.0086~a.u., respectively.
The difference between the RPA and DF results is taken to be the
uncertainty.
 Uncertainties  of the core and valence polarizabilities are added in quadrature to obtain uncertainties of the final values.

The final results listed in row labelled ``Total $\alpha_0 $''  are
compared with other theory~\cite{Mit08,HamHib08} and
experiment~\cite{KomGeaFeh05,Mit08}. Our value for the ground state
polarizability is in excellent agreement with both original RESIS
value ~\cite{KomGeaFeh05} value and the revised RESIS analysis
~\cite{Mit08}. Our values for the ground and $3s3p\ ^1P_1$ state
polarizabilities  are in excellent agreement with theoretical values
obtained with large-scale CI calculation with semiempirical
inclusion of the core polarization~\cite{Mit08}. The CI result of
\cite{HamHib08} is consistent with other  values; the small
difference is probably due to omission of the highly-excited states
in the valence CI and restricted treatment of the core excitations
in \cite{HamHib08}.

We note that the values of the $^1S_0$ and $^3P_0$ polarizabilities given in Table~\ref{table3} are very similar, their difference is only 10\% of
the ground state polarizability.

In summary, we have carried out a benchmark test of the theoretical and experimental determination of the ground state polarizability of the
Si$^{2+}$
 ion. Our final result is in excellent agreement with the RESIS experimental value ~\cite{KomGeaFeh05,Mit08}. High-precision
 recommended values are provided for the
  excited state $3s3p~^3P_0$ and $3s3p~^1P_1$ polarizabilities. The near cancellation of the $ns^2~^1S_0$ ground
state and the lowest $nsnp~^3P_0$ polarizabilities reported for B$^+$, Al$^+$, In$^+$, Tl$^+$, and Pb$^{2+}$ is also observed for the Si$^{2+}$ ion.

\section*{ACKNOWLEDGEMENTS}
We thank J. Mitroy for bringing our attention to this problem.  This
research was performed in part under the sponsorship of the US
Department of Commerce, National Institute of Standards and
Technology, and was supported by the National Science Foundation
under Physics Frontiers Center Grant PHY-0822671.
 The work of SGP was supported in part by US
NSF Grants No.\ PHY-1068699 and No.\ PHY-0758088. The work of MGK was
supported in part by RFBR grant No.\ 11-02-00943.

%\bibliography{bibfile2012}

\begin{thebibliography}{41}
\expandafter\ifx\csname natexlab\endcsname\relax\def\natexlab#1{#1}\fi \expandafter\ifx\csname bibnamefont\endcsname\relax
  \def\bibnamefont#1{#1}\fi
\expandafter\ifx\csname bibfnamefont\endcsname\relax
  \def\bibfnamefont#1{#1}\fi
\expandafter\ifx\csname citenamefont\endcsname\relax
  \def\citenamefont#1{#1}\fi
\expandafter\ifx\csname url\endcsname\relax
  \def\url#1{\texttt{#1}}\fi
\expandafter\ifx\csname urlprefix\endcsname\relax\def\urlprefix{URL }\fi \providecommand{\bibinfo}[2]{#2} \providecommand{\eprint}[2][]{\url{#2}}

\bibitem[{\citenamefont{{Mitroy} et~al.}(2010)\citenamefont{{Mitroy},
  {Safronova}, and {Clark}}}]{MitSafCla10}
\bibinfo{author}{\bibfnamefont{J.}~\bibnamefont{{Mitroy}}},
  \bibinfo{author}{\bibfnamefont{M.~S.} \bibnamefont{{Safronova}}},
  \bibnamefont{and} \bibinfo{author}{\bibfnamefont{C.~W.}
  \bibnamefont{{Clark}}}, \bibinfo{journal}{J. Phys. B}
  \textbf{\bibinfo{volume}{43}}, \bibinfo{pages}{202001}
  (\bibinfo{year}{2010}).

\bibitem[{\citenamefont{{Safronova} et~al.}(2010)\citenamefont{{Safronova},
  Jiang, Arora, Clark, Kozlov, Safronova, and Johnson}}]{SafJiaAro10}
\bibinfo{author}{\bibfnamefont{M.~S.} \bibnamefont{{Safronova}}},
  \bibinfo{author}{\bibfnamefont{D.}~\bibnamefont{Jiang}},
  \bibinfo{author}{\bibfnamefont{B.}~\bibnamefont{Arora}},
  \bibinfo{author}{\bibfnamefont{C.~W.} \bibnamefont{Clark}},
  \bibinfo{author}{\bibfnamefont{M.~G.} \bibnamefont{Kozlov}},
  \bibinfo{author}{\bibfnamefont{U.~I.} \bibnamefont{Safronova}},
  \bibnamefont{and} \bibinfo{author}{\bibfnamefont{W.~R.}
  \bibnamefont{Johnson}}, \bibinfo{journal}{IEEE Trans. Ultrason.
  Ferroelectrics and Frequency Control} \textbf{\bibinfo{volume}{57}},
  \bibinfo{pages}{94} (\bibinfo{year}{2010}).

\bibitem[{\citenamefont{Ekstrom et~al.}(1995)\citenamefont{Ekstrom,
  Schmiedmayer, Chapman, Hammond, and Pritchard}}]{EksSchCha95}
\bibinfo{author}{\bibfnamefont{C.~R.} \bibnamefont{Ekstrom}},
  \bibinfo{author}{\bibfnamefont{J.}~\bibnamefont{Schmiedmayer}},
  \bibinfo{author}{\bibfnamefont{M.~S.} \bibnamefont{Chapman}},
  \bibinfo{author}{\bibfnamefont{T.~D.} \bibnamefont{Hammond}},
  \bibnamefont{and} \bibinfo{author}{\bibfnamefont{D.~E.}
  \bibnamefont{Pritchard}}, \bibinfo{journal}{Phys. Rev. A}
  \textbf{\bibinfo{volume}{51}}, \bibinfo{pages}{3883} (\bibinfo{year}{1995}).

\bibitem[{\citenamefont{Derevianko et~al.}(1999)\citenamefont{Derevianko,
  Johnson, Safronova, and Babb}}]{DerJohSaf99}
\bibinfo{author}{\bibfnamefont{A.}~\bibnamefont{Derevianko}},
  \bibinfo{author}{\bibfnamefont{W.~R.} \bibnamefont{Johnson}},
  \bibinfo{author}{\bibfnamefont{M.~S.} \bibnamefont{Safronova}},
  \bibnamefont{and} \bibinfo{author}{\bibfnamefont{J.~F.} \bibnamefont{Babb}},
  \bibinfo{journal}{Phys. Rev. Lett.} \textbf{\bibinfo{volume}{82}},
  \bibinfo{pages}{3589} (\bibinfo{year}{1999}).

\bibitem[{\citenamefont{Amini and Gould}(2003)}]{AmiGou03}
\bibinfo{author}{\bibfnamefont{J.~M.} \bibnamefont{Amini}} \bibnamefont{and}
  \bibinfo{author}{\bibfnamefont{H.}~\bibnamefont{Gould}},
  \bibinfo{journal}{Phys. Rev. Lett.} \textbf{\bibinfo{volume}{91}},
  \bibinfo{pages}{153001 } (\bibinfo{year}{2003}).

\bibitem[{\citenamefont{Safronova and Clark}(2004)}]{SafCla04}
\bibinfo{author}{\bibfnamefont{M.~S.} \bibnamefont{Safronova}}
  \bibnamefont{and} \bibinfo{author}{\bibfnamefont{C.~W.} \bibnamefont{Clark}},
  \bibinfo{journal}{Phys. Rev. A} \textbf{\bibinfo{volume}{69}},
  \bibinfo{pages}{40501} (\bibinfo{year}{2004}).

\bibitem[{\citenamefont{Auzinsh et~al.}(2007)\citenamefont{Auzinsh, Bluss,
  Ferber, Gahbauer, Jarmola, Safronova, Safronova, and Tamanis}}]{AuzBluFer07}
\bibinfo{author}{\bibfnamefont{M.}~\bibnamefont{Auzinsh}},
  \bibinfo{author}{\bibfnamefont{K.}~\bibnamefont{Bluss}},
  \bibinfo{author}{\bibfnamefont{R.}~\bibnamefont{Ferber}},
  \bibinfo{author}{\bibfnamefont{F.}~\bibnamefont{Gahbauer}},
  \bibinfo{author}{\bibfnamefont{A.}~\bibnamefont{Jarmola}},
  \bibinfo{author}{\bibfnamefont{M.~S.} \bibnamefont{Safronova}},
  \bibinfo{author}{\bibfnamefont{U.~I.} \bibnamefont{Safronova}},
  \bibnamefont{and} \bibinfo{author}{\bibfnamefont{M.}~\bibnamefont{Tamanis}},
  \bibinfo{journal}{Phys. Rev. A} \textbf{\bibinfo{volume}{75}},
  \bibinfo{pages}{22502} (\bibinfo{year}{2007}).

\bibitem[{\citenamefont{Gunawardena et~al.}(2007)\citenamefont{Gunawardena,
  Elliott, Safronova, and Safronova}}]{GunEllSaf07}
\bibinfo{author}{\bibfnamefont{M.}~\bibnamefont{Gunawardena}},
  \bibinfo{author}{\bibfnamefont{D.~S.} \bibnamefont{Elliott}},
  \bibinfo{author}{\bibfnamefont{M.~S.} \bibnamefont{Safronova}},
  \bibnamefont{and} \bibinfo{author}{\bibfnamefont{U.~I.}
  \bibnamefont{Safronova}}, \bibinfo{journal}{Phys. Rev. A}
  \textbf{\bibinfo{volume}{75}}, \bibinfo{pages}{22507} (\bibinfo{year}{2007}).

\bibitem[{\citenamefont{Iskrenova-Tchoukova and Safronova}(2008)}]{IskSaf08}
\bibinfo{author}{\bibfnamefont{E.}~\bibnamefont{Iskrenova-Tchoukova}}
  \bibnamefont{and} \bibinfo{author}{\bibfnamefont{M.~S.}
  \bibnamefont{Safronova}}, \bibinfo{journal}{Phys. Rev. A}
  \textbf{\bibinfo{volume}{78}}, \bibinfo{pages}{012508}
  (\bibinfo{year}{2008}).

\bibitem[{\citenamefont{Mitroy and Safronova}(2009)}]{MitSaf09}
\bibinfo{author}{\bibfnamefont{J.}~\bibnamefont{Mitroy}} \bibnamefont{and}
  \bibinfo{author}{\bibfnamefont{M.~S.} \bibnamefont{Safronova}},
  \bibinfo{journal}{Phys. Rev. A} \textbf{\bibinfo{volume}{79}},
  \bibinfo{pages}{012513} (\bibinfo{year}{2009}).

\bibitem[{\citenamefont{{Holmgren} et~al.}(2010)\citenamefont{{Holmgren},
  {Revelle}, {Lonij}, and {Cronin}}}]{HolRevLon10}
\bibinfo{author}{\bibfnamefont{W.~F.} \bibnamefont{{Holmgren}}},
  \bibinfo{author}{\bibfnamefont{M.~C.} \bibnamefont{{Revelle}}},
  \bibinfo{author}{\bibfnamefont{V.~P.~A.} \bibnamefont{{Lonij}}},
  \bibnamefont{and} \bibinfo{author}{\bibfnamefont{A.~D.}
  \bibnamefont{{Cronin}}}, \bibinfo{journal}{\pra}
  \textbf{\bibinfo{volume}{81}}, \bibinfo{eid}{053607} (\bibinfo{year}{2010}).

\bibitem[{\citenamefont{Kortyna et~al.}(2011)\citenamefont{Kortyna, Tinsman,
  Grab, Safronova, and Safronova}}]{KorTinGra11}
\bibinfo{author}{\bibfnamefont{A.}~\bibnamefont{Kortyna}},
  \bibinfo{author}{\bibfnamefont{C.}~\bibnamefont{Tinsman}},
  \bibinfo{author}{\bibfnamefont{J.}~\bibnamefont{Grab}},
  \bibinfo{author}{\bibfnamefont{M.~S.} \bibnamefont{Safronova}},
  \bibnamefont{and} \bibinfo{author}{\bibfnamefont{U.~I.}
  \bibnamefont{Safronova}}, \bibinfo{journal}{Phys. Rev. A}
  \textbf{\bibinfo{volume}{83}}, \bibinfo{pages}{042511}
  (\bibinfo{year}{2011}).

\bibitem[{\citenamefont{Porsev et~al.}(2008)\citenamefont{Porsev, Ludlow, Boyd,
  and Ye}}]{PorLudBoy08}
\bibinfo{author}{\bibfnamefont{S.~G.} \bibnamefont{Porsev}},
  \bibinfo{author}{\bibfnamefont{A.~D.} \bibnamefont{Ludlow}},
  \bibinfo{author}{\bibfnamefont{M.~M.} \bibnamefont{Boyd}}, \bibnamefont{and}
  \bibinfo{author}{\bibfnamefont{J.}~\bibnamefont{Ye}}, \bibinfo{journal}{Phys.
  Rev. A} \textbf{\bibinfo{volume}{78}}, \bibinfo{pages}{032508}
  (\bibinfo{year}{2008}).

\bibitem[{\citenamefont{{Rosenband} et~al.}(2008)}]{RosHumSch08}
\bibinfo{author}{\bibfnamefont{T.}~\bibnamefont{{Rosenband}}}
  \bibnamefont{et~al.}, \bibinfo{journal}{Science}
  \textbf{\bibinfo{volume}{319}}, \bibinfo{pages}{1808} (\bibinfo{year}{2008}).

\bibitem[{\citenamefont{Jiang et~al.}(2011)\citenamefont{Jiang, Ludlow, Lemke,
  Fox, Sherman, Ma, and Oates}}]{JiaLudLem11}
\bibinfo{author}{\bibfnamefont{Y.~Y.} \bibnamefont{Jiang}},
  \bibinfo{author}{\bibfnamefont{A.~D.} \bibnamefont{Ludlow}},
  \bibinfo{author}{\bibfnamefont{N.~D.} \bibnamefont{Lemke}},
  \bibinfo{author}{\bibfnamefont{R.~W.} \bibnamefont{Fox}},
  \bibinfo{author}{\bibfnamefont{J.~A.} \bibnamefont{Sherman}},
  \bibinfo{author}{\bibfnamefont{L.-S.} \bibnamefont{Ma}}, \bibnamefont{and}
  \bibinfo{author}{\bibfnamefont{C.~W.} \bibnamefont{Oates}},
  \bibinfo{journal}{Nat. Photonics} \textbf{\bibinfo{volume}{5}},
  \bibinfo{pages}{158 } (\bibinfo{year}{2011}).

\bibitem[{\citenamefont{Hachisu et~al.}(2008)\citenamefont{Hachisu, Miyagishi,
  Porsev, Derevianko, Ovsiannikov, Palchikov, Takamoto, and
  Katori}}]{HacMiyPor08}
\bibinfo{author}{\bibfnamefont{H.}~\bibnamefont{Hachisu}},
  \bibinfo{author}{\bibfnamefont{K.}~\bibnamefont{Miyagishi}},
  \bibinfo{author}{\bibfnamefont{S.~G.} \bibnamefont{Porsev}},
  \bibinfo{author}{\bibfnamefont{A.}~\bibnamefont{Derevianko}},
  \bibinfo{author}{\bibfnamefont{V.~D.} \bibnamefont{Ovsiannikov}},
  \bibinfo{author}{\bibfnamefont{V.~G.} \bibnamefont{Palchikov}},
  \bibinfo{author}{\bibfnamefont{M.}~\bibnamefont{Takamoto}}, \bibnamefont{and}
  \bibinfo{author}{\bibfnamefont{H.}~\bibnamefont{Katori}},
  \bibinfo{journal}{Phys. Rev. Lett.} \textbf{\bibinfo{volume}{100}},
  \bibinfo{pages}{053001} (\bibinfo{year}{2008}).

\bibitem[{\citenamefont{Gorshkov et~al.}(2009)\citenamefont{Gorshkov, Rey,
  Daley, Boyd, Ye, Zoller, and Lukin}}]{GorReyDal09}
\bibinfo{author}{\bibfnamefont{A.~V.} \bibnamefont{Gorshkov}},
  \bibinfo{author}{\bibfnamefont{A.~M.} \bibnamefont{Rey}},
  \bibinfo{author}{\bibfnamefont{A.~J.} \bibnamefont{Daley}},
  \bibinfo{author}{\bibfnamefont{M.~M.} \bibnamefont{Boyd}},
  \bibinfo{author}{\bibfnamefont{J.}~\bibnamefont{Ye}},
  \bibinfo{author}{\bibfnamefont{P.}~\bibnamefont{Zoller}}, \bibnamefont{and}
  \bibinfo{author}{\bibfnamefont{M.~D.} \bibnamefont{Lukin}},
  \bibinfo{journal}{Phys. Rev. Lett.} \textbf{\bibinfo{volume}{102}},
  \bibinfo{pages}{110503} (\bibinfo{year}{2009}).

\bibitem[{\citenamefont{Maroulis}(2006)}]{Mar06}
\bibinfo{editor}{\bibfnamefont{G.}~\bibnamefont{Maroulis}}, ed.,
  \emph{\bibinfo{title}{Atoms, Molecules and Clusters in Electric Fields.
  Theoretical Approaches to the Calculation of Electric Polarizability}}
  (\bibinfo{publisher}{Imperial College Press}, \bibinfo{year}{2006}).

\bibitem[{\citenamefont{{Ward} et~al.}(1996)\citenamefont{{Ward}, {Sturrus},
  and {Lundeen}}}]{WarStuLun96}
\bibinfo{author}{\bibfnamefont{R.~F.} \bibnamefont{{Ward}},
  \bibfnamefont{Jr.}}, \bibinfo{author}{\bibfnamefont{W.~G.}
  \bibnamefont{{Sturrus}}}, \bibnamefont{and}
  \bibinfo{author}{\bibfnamefont{S.~R.} \bibnamefont{{Lundeen}}},
  \bibinfo{journal}{\pra} \textbf{\bibinfo{volume}{53}}, \bibinfo{pages}{113}
  (\bibinfo{year}{1996}).

\bibitem[{\citenamefont{{Komara} et~al.}(2005)\citenamefont{{Komara}, {Gearba},
  {Fehrenbach}, and {Lundeen}}}]{KomGeaFeh05}
\bibinfo{author}{\bibfnamefont{R.~A.} \bibnamefont{{Komara}}},
  \bibinfo{author}{\bibfnamefont{M.~A.} \bibnamefont{{Gearba}}},
  \bibinfo{author}{\bibfnamefont{C.~W.} \bibnamefont{{Fehrenbach}}},
  \bibnamefont{and} \bibinfo{author}{\bibfnamefont{S.~R.}
  \bibnamefont{{Lundeen}}}, \bibinfo{journal}{J. Phys. B}
  \textbf{\bibinfo{volume}{38}}, \bibinfo{pages}{87} (\bibinfo{year}{2005}).

\bibitem[{\citenamefont{{Jacobson} et~al.}(2000)\citenamefont{{Jacobson},
  {Komara}, {Sturrus}, and {Lundeen}}}]{JacKomStu00}
\bibinfo{author}{\bibfnamefont{P.~L.} \bibnamefont{{Jacobson}}},
  \bibinfo{author}{\bibfnamefont{R.~A.} \bibnamefont{{Komara}}},
  \bibinfo{author}{\bibfnamefont{W.~G.} \bibnamefont{{Sturrus}}},
  \bibnamefont{and} \bibinfo{author}{\bibfnamefont{S.~R.}
  \bibnamefont{{Lundeen}}}, \bibinfo{journal}{\pra}
  \textbf{\bibinfo{volume}{62}}, \bibinfo{eid}{012509} (\bibinfo{year}{2000}).

\bibitem[{\citenamefont{{Snow} and {Lundeen}}(2008)}]{SnoLun08}
\bibinfo{author}{\bibfnamefont{E.~L.} \bibnamefont{{Snow}}} \bibnamefont{and}
  \bibinfo{author}{\bibfnamefont{S.~R.} \bibnamefont{{Lundeen}}},
  \bibinfo{journal}{\pra} \textbf{\bibinfo{volume}{77}}, \bibinfo{eid}{052501}
  (\bibinfo{year}{2008}).

\bibitem[{\citenamefont{{Komara} et~al.}(2003)\citenamefont{{Komara}, {Gearba},
  {Lundeen}, and {Fehrenbach}}}]{KomGeaLun03}
\bibinfo{author}{\bibfnamefont{R.~A.} \bibnamefont{{Komara}}},
  \bibinfo{author}{\bibfnamefont{M.~A.} \bibnamefont{{Gearba}}},
  \bibinfo{author}{\bibfnamefont{S.~R.} \bibnamefont{{Lundeen}}},
  \bibnamefont{and} \bibinfo{author}{\bibfnamefont{C.~W.}
  \bibnamefont{{Fehrenbach}}}, \bibinfo{journal}{\pra}
  \textbf{\bibinfo{volume}{67}}, \bibinfo{eid}{062502} (\bibinfo{year}{2003}).

\bibitem[{\citenamefont{{Snow} et~al.}(2005)\citenamefont{{Snow}, {Gearba},
  {Komara}, {Lundeen}, and {Sturrus}}}]{SnoGeaKom05}
\bibinfo{author}{\bibfnamefont{E.~L.} \bibnamefont{{Snow}}},
  \bibinfo{author}{\bibfnamefont{M.~A.} \bibnamefont{{Gearba}}},
  \bibinfo{author}{\bibfnamefont{R.~A.} \bibnamefont{{Komara}}},
  \bibinfo{author}{\bibfnamefont{S.~R.} \bibnamefont{{Lundeen}}},
  \bibnamefont{and} \bibinfo{author}{\bibfnamefont{W.~G.}
  \bibnamefont{{Sturrus}}}, \bibinfo{journal}{\pra}
  \textbf{\bibinfo{volume}{71}}, \bibinfo{eid}{022510} (\bibinfo{year}{2005}).

\bibitem[{\citenamefont{{Hanni} et~al.}(2010)\citenamefont{{Hanni}, {Keele},
  {Lundeen}, {Fehrenbach}, and {Sturrus}}}]{HanKeeLun10}
\bibinfo{author}{\bibfnamefont{M.~E.} \bibnamefont{{Hanni}}},
  \bibinfo{author}{\bibfnamefont{J.~A.} \bibnamefont{{Keele}}},
  \bibinfo{author}{\bibfnamefont{S.~R.} \bibnamefont{{Lundeen}}},
  \bibinfo{author}{\bibfnamefont{C.~W.} \bibnamefont{{Fehrenbach}}},
  \bibnamefont{and} \bibinfo{author}{\bibfnamefont{W.~G.}
  \bibnamefont{{Sturrus}}}, \bibinfo{journal}{\pra}
  \textbf{\bibinfo{volume}{81}}, \bibinfo{eid}{042512} (\bibinfo{year}{2010}).

\bibitem[{\citenamefont{{Keele}
  et~al.}(2011{\natexlab{a}})\citenamefont{{Keele}, {Hanni}, {Woods},
  {Lundeen}, and {Fehrenbach}}}]{KeeHanWoo11}
\bibinfo{author}{\bibfnamefont{J.~A.} \bibnamefont{{Keele}}},
  \bibinfo{author}{\bibfnamefont{M.~E.} \bibnamefont{{Hanni}}},
  \bibinfo{author}{\bibfnamefont{S.~L.} \bibnamefont{{Woods}}},
  \bibinfo{author}{\bibfnamefont{S.~R.} \bibnamefont{{Lundeen}}},
  \bibnamefont{and} \bibinfo{author}{\bibfnamefont{C.~W.}
  \bibnamefont{{Fehrenbach}}}, \bibinfo{journal}{Phys. Rev. A}
  \textbf{\bibinfo{volume}{83}}, \bibinfo{eid}{062501}
  (\bibinfo{year}{2011}{\natexlab{a}}).

\bibitem[{\citenamefont{{Keele}
  et~al.}(2011{\natexlab{b}})\citenamefont{{Keele}, {Lundeen}, and
  {Fehrenbach}}}]{KeeLunFeh11}
\bibinfo{author}{\bibfnamefont{J.~A.} \bibnamefont{{Keele}}},
  \bibinfo{author}{\bibfnamefont{S.~R.} \bibnamefont{{Lundeen}}},
  \bibnamefont{and} \bibinfo{author}{\bibfnamefont{C.~W.}
  \bibnamefont{{Fehrenbach}}}, \bibinfo{journal}{Phys. Rev. A}
  \textbf{\bibinfo{volume}{83}}, \bibinfo{eid}{062509}
  (\bibinfo{year}{2011}{\natexlab{b}}).

\bibitem[{\citenamefont{{Lundeen} and {Fehrenbach}}(2007)}]{LunFeh07}
\bibinfo{author}{\bibfnamefont{S.~R.} \bibnamefont{{Lundeen}}}
  \bibnamefont{and} \bibinfo{author}{\bibfnamefont{C.~W.}
  \bibnamefont{{Fehrenbach}}}, \bibinfo{journal}{\pra}
  \textbf{\bibinfo{volume}{75}}, \bibinfo{eid}{032523} (\bibinfo{year}{2007}).

\bibitem[{\citenamefont{{Mitroy}}(2008)}]{Mit08}
\bibinfo{author}{\bibfnamefont{J.}~\bibnamefont{{Mitroy}}},
  \bibinfo{journal}{Phys. Rev. A} \textbf{\bibinfo{volume}{78}},
  \bibinfo{eid}{052515} (\bibinfo{year}{2008}).

\bibitem[{\citenamefont{{Safronova} et~al.}(2011)\citenamefont{{Safronova},
  {Kozlov}, and {Clark}}}]{SafKozCla11}
\bibinfo{author}{\bibfnamefont{M.~S.} \bibnamefont{{Safronova}}},
  \bibinfo{author}{\bibfnamefont{M.~G.} \bibnamefont{{Kozlov}}},
  \bibnamefont{and} \bibinfo{author}{\bibfnamefont{C.~W.}
  \bibnamefont{{Clark}}}, \bibinfo{journal}{Phys. Rev. Lett.}
  \textbf{\bibinfo{volume}{107}}, \bibinfo{eid}{143006} (\bibinfo{year}{2011}).

\bibitem[{\citenamefont{{Safronova} et~al.}(2009)\citenamefont{{Safronova},
  {Kozlov}, {Johnson}, and {Jiang}}}]{SafKozJoh09}
\bibinfo{author}{\bibfnamefont{M.~S.} \bibnamefont{{Safronova}}},
  \bibinfo{author}{\bibfnamefont{M.~G.} \bibnamefont{{Kozlov}}},
  \bibinfo{author}{\bibfnamefont{W.~R.} \bibnamefont{{Johnson}}},
  \bibnamefont{and} \bibinfo{author}{\bibfnamefont{D.}~\bibnamefont{{Jiang}}},
  \bibinfo{journal}{Phys. Rev. A} \textbf{\bibinfo{volume}{80}},
  \bibinfo{eid}{012516} (\bibinfo{year}{2009}).

\bibitem[{\citenamefont{Kozlov}(2004)}]{Koz04}
\bibinfo{author}{\bibfnamefont{M.~G.} \bibnamefont{Kozlov}},
  \bibinfo{journal}{Int. J. Quant. Chem.} \textbf{\bibinfo{volume}{100}},
  \bibinfo{pages}{336} (\bibinfo{year}{2004}).

\bibitem[{\citenamefont{{Safronova} et~al.}(2012)\citenamefont{{Safronova},
  {Kozlov}, and {Safronova}}}]{SafKozSaf12}
\bibinfo{author}{\bibfnamefont{M.~S.} \bibnamefont{{Safronova}}},
  \bibinfo{author}{\bibfnamefont{M.~G.} \bibnamefont{{Kozlov}}},
  \bibnamefont{and} \bibinfo{author}{\bibfnamefont{U.~I.}
  \bibnamefont{{Safronova}}}, \bibinfo{journal}{\pra}
  \textbf{\bibinfo{volume}{85}}, \bibinfo{eid}{012507} (\bibinfo{year}{2012}).

\bibitem[{Ral()}]{RalKraRea11}
\bibinfo{note}{Yu.~Ralchenko, A.~Kramida, J.~Reader, and NIST ASD Team (2011).
  NIST Atomic Spectra Database (version 4.1), [Online]. Available:
  http://physics.nist.gov/asd. National Institute of Standards and Technology,
  Gaithersburg, MD.}

\bibitem[{\citenamefont{Chou et~al.}(2010)\citenamefont{Chou, Hume, Koelemeij,
  Wineland, and Rosenband}}]{ChoHumKoe10}
\bibinfo{author}{\bibfnamefont{C.~W.} \bibnamefont{Chou}},
  \bibinfo{author}{\bibfnamefont{D.~B.} \bibnamefont{Hume}},
  \bibinfo{author}{\bibfnamefont{J.~C.~J.} \bibnamefont{Koelemeij}},
  \bibinfo{author}{\bibfnamefont{D.~J.} \bibnamefont{Wineland}},
  \bibnamefont{and}
  \bibinfo{author}{\bibfnamefont{T.}~\bibnamefont{Rosenband}},
  \bibinfo{journal}{Phys. Rev. Lett.} p. \bibinfo{pages}{070802}
  (\bibinfo{year}{2010}).

\bibitem[{\citenamefont{Porsev and Derevianko}(2006)}]{PorDer06}
\bibinfo{author}{\bibfnamefont{S.~G.} \bibnamefont{Porsev}} \bibnamefont{and}
  \bibinfo{author}{\bibfnamefont{A.}~\bibnamefont{Derevianko}},
  \bibinfo{journal}{Phys. Rev. A} \textbf{\bibinfo{volume}{74}},
  \bibinfo{pages}{020502(R)} (\bibinfo{year}{2006}).

\bibitem[{\citenamefont{Kotochigova and Tupitsyn}(1987)}]{KotTup87}
\bibinfo{author}{\bibfnamefont{S.~A.} \bibnamefont{Kotochigova}}
  \bibnamefont{and} \bibinfo{author}{\bibfnamefont{I.~I.}
  \bibnamefont{Tupitsyn}}, \bibinfo{journal}{J. Phys. B}
  \textbf{\bibinfo{volume}{20}}, \bibinfo{pages}{4759} (\bibinfo{year}{1987}).

\bibitem[{\citenamefont{Dzuba et~al.}(1996)\citenamefont{Dzuba, Flambaum, and
  Kozlov}}]{DzuFlaKoz96b}
\bibinfo{author}{\bibfnamefont{V.~A.} \bibnamefont{Dzuba}},
  \bibinfo{author}{\bibfnamefont{V.~V.} \bibnamefont{Flambaum}},
  \bibnamefont{and} \bibinfo{author}{\bibfnamefont{M.~G.}
  \bibnamefont{Kozlov}}, \bibinfo{journal}{Phys.\ Rev.\ A}
  \textbf{\bibinfo{volume}{54}}, \bibinfo{pages}{3948} (\bibinfo{year}{1996}).

\bibitem[{Koz()}]{KozPorTup00}
\bibinfo{note}{{M}. G. Kozlov, S. G. Porsev, and I. I. Tupitsyn,
  arXiv:physics/0004076 (2000)}.

\bibitem[{\citenamefont{{Kozlov} and {Porsev}}(1999)}]{KozPor99a}
\bibinfo{author}{\bibfnamefont{M.~G.} \bibnamefont{{Kozlov}}} \bibnamefont{and}
  \bibinfo{author}{\bibfnamefont{S.~G.} \bibnamefont{{Porsev}}},
  \bibinfo{journal}{Eur.~Phys.~J.~D} \textbf{\bibinfo{volume}{5}},
  \bibinfo{pages}{59} (\bibinfo{year}{1999}).

\bibitem[{\citenamefont{{Hamonou} and {Hibbert}}(2008)}]{HamHib08}
\bibinfo{author}{\bibfnamefont{L.}~\bibnamefont{{Hamonou}}} \bibnamefont{and}
  \bibinfo{author}{\bibfnamefont{A.}~\bibnamefont{{Hibbert}}},
  \bibinfo{journal}{J. Phys. B} \textbf{\bibinfo{volume}{41}},
  \bibinfo{pages}{245004} (\bibinfo{year}{2008}).

\end{thebibliography}

\end{document}